\documentclass[12pt,preprint]{aastex}

\usepackage{amsmath}

\slugcomment{KSUPT -- 08/3  \quad  March 2009}

\begin{document}

\title{Constraints on dark energy from baryon acoustic peak and galaxy cluster gas mass measurements}

\author{Lado Samushia\altaffilmark{1,2} and Bharat Ratra\altaffilmark{1}}

\altaffiltext{1}{Department of Physics, Kansas State University, 116 Cardwell Hall, Manhattan, KS 66506, lado@phys.ksu.edu, ratra@phys.ksu.edu.}
\altaffiltext{2}{National Abastumani Astrophysical Observatory, 2A Kazbegi Ave, GE-0160 Tbilisi, Georgia}

\begin{abstract}
We use baryon acoustic peak measurements by \citet{eisensteinetal} and
\citet{percivaletal07a}, together with the WMAP measurement of the apparent acoustic horizon angle, and galaxy cluster gas mass fraction
measurements of \citet{allen08}, to constrain a slowly rolling scalar field dark energy model, $\phi$CDM, in which dark energy's energy density changes in time. We also compare our $\phi$CDM results with those derived for two more common dark energy models: the time-independent cosmological constant model, $\Lambda$CDM, and the XCDM parametrization of dark energy's equation of state. For time-independent dark energy, the \citet{percivaletal07a} measurements effectively constrain spatial curvature and favor a close to spatially-flat model, mostly due to the WMAP CMB prior used in the analysis. In a spatially-flat model the \citet{percivaletal07a} data less effectively constrain time-varying dark energy. The joint baryon acoustic peak and galaxy cluster gas mass constraints on $\phi$CDM model are consistent with but tighter than those derived from other data. A time-independent cosmological constant in a spatially-flat model provides a good fit to the joint data, while the $\alpha$ parameter in the inverse power law potential $\phi$CDM model is constrained to be less than about 4 at 3$\sigma$ confidence level.
\end{abstract}


\section{Introduction}

About a decade ago type Ia supernova (SNIa) observations provided initial evidence that the cosmological expansion is accelerating \citep{riess98, perlmutter99}. If general relativity is valid on the scales of current cosmological observations, more recent SNIa data as well as results of various other cosmological tests including large-scale structure tests and observations of the cosmic microwave background (CMB) anisotropy can be reasonably well reconciled if we assume that about two-thirds of the cosmological energy budget is in the form of dark energy.

Different theoretical models of dark energy have been proposed over the years. A few models try to do away with the need for an exotic dark energy component by modifying general relativity on large scales \citep[see, e.g.,][]{wang07a, demianski07, tsujikawa08, capozziello08, wei08, gannouji08}. If general relativity is valid we need a substance that has negative pressure, $p<-\rho/3$ (where $\rho$ is the energy density), to have accelerated cosmological expansion. The simplest standard cosmological model is $\Lambda$CDM \citep{pee84} in which the cosmological constant $\Lambda$ has negative pressure and powers the current accelerated expansion of the universe. Although $\Lambda$ has a quantum field theory motivation as vacuum energy, $\Lambda$CDM has a number of apparent problems. The most celebrated is the fact that the value of vacuum energy density calculated from field theory with a Planck scale cutoff is many orders of magnitude larger than the measured value. Because of this other models have been developed, despite the fact that the simple $\Lambda$CDM model provides a fairly good fit to most cosmological data. In our paper we also study the slowly rolling scalar field dark energy model  \citep[$\phi$CDM,][]{pee88, rat88}. In the $\phi$CDM model the small (classical) value of the current vacuum energy density is a consequence of the scalar field dynamics. The third model we consider is the XCDM parametrization. XCDM parametrizes dark energy's equation of state as $p_{\rm x}=\omega_{\rm x}\rho_{\rm x}$, where $\omega_{\rm x}$ is a negative constant. This approximation is not accurate in the scalar field dominated epoch \citep{rat91}.\footnote{For recent reviews of dark energy see, e.g., \citet{rat08}, \citet{linder08}, \citet{frieman08}, and \citet{martin08}.}$^,$\footnote{We assume that dark energy and dark matter only couple gravitationally. For discussion of models with other couplings see, e.g., \citet{costa08}, \citet{mainini07}, \citet{brookfield08}, \citet{he08}, and \citet{olivares08}. For other models of dark energy see, e.g., \citet{grande07}, \citet{neupane08}, \citet{mathews08}, \citet{usmani08}, and \citet{ichiki08}.}

In the $\phi$CDM model one can explain the accelerated expansion of the universe by introducing a scalar field $\phi$ minimally coupled to gravity. The action for such a term is

\begin{equation}
S_{\phi}=\int d^4x\sqrt{-g}\left[\frac{1}{2}g^{\mu\nu}\partial_{\mu}\phi\partial_{\nu}\phi-\frac{V(\phi)}{G}\right],
\end{equation}

\noindent
where $G$ is the gravitational constant. If the scalar field is close to homogeneous on cosmological scales, then to leading order it's energy density and pressure are given by

\begin{equation}
\rho_{\phi}=\frac{1}{2}\left(\frac{d\phi}{dt}\right)^2+\frac{V(\phi)}{G},
\end{equation}

\begin{equation}
p_\phi=\frac{1}{2}\left(\frac{d\phi}{dt}\right)^2-\frac{V(\phi)}{G},
\end{equation}

\noindent
When the scalar field changes only slowly in time, the effective equation of state parameter $\omega_{\phi}=p_{\phi}/\rho_{\phi}$ is negative and the scalar field acts like a time-dependent cosmological constant. To specify the $\phi$CDM model one has to pick a specific form of potential energy density $V(\phi)$. Neither cosmological observations nor fundamental particle physics theory can provide significant motivation for a specific form of potential energy and a lot of different cases have been studied. In our paper we work with the inverse power law potential energy density $V(\phi)\varpropto \phi^{-\alpha}$, because it has been well studied and it provides a practical way of parametrizing the slowly rolling scalar field with one nonnegative dimensionless parameter $\alpha$. Physically, large values of $\alpha$ correspond to rapid time evolution, while the limit of $\alpha=0$ gives a time-independent cosmological constant. \citet[Fig.\ 2]{podariu00} relate this $\phi$CDM model to the XCDM parametrization and discuss how $\alpha$ and effective $\omega_{\phi}$ are related. For large values of $\alpha$ the time-dependent equation of state parameter changes very fast and the XCDM parametrization fails to provide a good phenomenological description of the scalar field. Figure~1 shows the residuals between comoving distance calculated in $\phi$CDM and XCDM models. Already for $\alpha=2.0$ the predictions of XCDM differ significantly at high redshifts. If $\alpha$ is very close to zero the scalar field equation of state changes very slowly and becomes more and more difficult to distinguish from $\Lambda$CDM and the XCDM parametrization is reasonable. If $\alpha$ turns out to be a very small but nonzero number, a lot of independent high precision cosmological measurements supplemented with the better understanding of underlying high energy physics will be necessary to discriminate between different dark energy models.

Assuming the cold dark matter (CDM) model of structure formation \citep[for a discussion of apparent problems with this model see][and references therein]{pee03}, and assuming that the dark energy is a time-independent cosmological constant \citep[see, e.g.,][]{wang07, gong07, ichikawa08, virey08}, CMB anisotropy data combined with independent dark matter density measurements \citep[see, e.g.,][]{che03b} are consistent with negligible spatial curvature \citep[see, e.g.,][]{pod01b, page03, spergel07, doran07}. CMB anisotropy data in combination with the low measured density of nonrelativistic matter then require the presence of dark energy and so are consistent with the SNIa results.

Many different observational tests have been used to constrain cosmological parameters. An issue of great current interest is whether dark energy is Einstein's cosmological constant or whether it evolves slowly in time and varies weakly in space. Current SNIa data are unable to resolve this \citep[see, e.g.,][]{mignone07, wuyu08, lin08, dev08, liu08, kowalski08}, but future SNIa data will improve the constraints \citep[see, e.g.,][]{pod01a} and, unless $\alpha$ has a very small value, might be able to detect time variation of dark energy. Current SNIa and CMB data are consistent with the $\Lambda$CDM model, but it is not yet possible to reject other dark energy models with high statistical confidence \citep[see, e.g.,][]{rap05, wilson06, davis07}.

To tighten the constraints on cosmological parameters, it is important to have many independent tests of dark energy models. Comparison of constraints from different tests can help uncover unknown systematic effects, and combinations of constraints from different tests can better discriminate between models. Other observational tests under recent discussion include the angular size of radio sources and quasars as a function of redshift \citep[see, e.g.,][]{chen03a, pod03, dal07, santos08}, strong gravitational lensing \citep[see, e.g.,][]{lee07, oguri08, zhang07, zhu08}, weak gravitational lensing \citep[see, e.g.,][]{takada08, fu07, dore07, lavacca08}, measurements of the Hubble parameter as a function of redshift \citep[see, e.g.,][]{samushia06, lazkoz07b, weizhang08, szydlowski08}, large-scale structure baryon acoustic oscillation peak measurements \citep[see, e.g.,][]{xia07, lim07, sapone07} and galaxy cluster gas mass fraction versus redshift data \citep[see, e.g.,][]{allen04, che04, sen08}. For recent reviews of the observational constraints on dark energy see, e.g., \citet{kur07} and \citet{wang07b}.

Many different observational test have been used to constrain the slowly-rolling scalar field dark energy model. The constraints are getting tighter as the quality and quantity of new measurements is increasing. Constraints on the $\alpha$ parameter from different tests are shown in Table\ 1. In our paper we use baryon acoustic oscillations (BAO) peak measurements to constrain the $\phi$CDM model of dark energy and compare our results to the constraints on $\Lambda$CDM model and XCDM parametrization. Since the peak has been measured at only two redshifts, $z=0.2$ and $z=0.35$ \citep{eisensteinetal, percivaletal07a}, BAO data alone can not tightly constrain the models. To more tightly constrain the dark energy models, we perform a joint analysis of the BAO data with new galaxy cluster gas mass fraction versus redshift data \citep{allen08}.\footnote{The galaxy cluster gas mass fraction test was proposed by \citet{sasaki96} and \citet{pen97}.} The resulting constraints are consistent with, but typically more constraining than, those derived from other data (see Table\ 1).

In Sec.\ 2 we briefly describe the BAO method we use. In Sec.\ 3 we summarize the BAO and galaxy cluster gas mass fraction data and computations. We discuss our results in Sec.\ 4.

\section{Baryon acoustic oscillations}

Before recombination baryons and photons are tightly coupled and gravity and pressure gradients induce sub-acoustic-Hubble-radius oscillations in the baryon-photon fluid \citep{sunyaev70, peebles70}. These transmute into the acoustic peaks observed now in the CMB anisotropy angular power spectrum, which provide very useful information on various cosmological parameters. The baryonic matter gravitationally interacts with the dark matter and so the matter power spectrum should also exhibit these ``baryon acoustic" wiggles. Because the baryonic matter is a small fraction of the total matter the amplitudes of the BAO wiggles are small. The BAO peak length scale is set by the sound horizon at decoupling, $\sim 10^2 \ {\rm Mpc}$, and so detecting the BAO peak in a real space correlation function requires observationally sampling a large volume. The BAO peak in the galaxy correlation function has recently been detected by using SDSS data (\citealt{eisensteinetal}, also see \citealt{hutsi06}) and by using 2dFGRS data \citep{cole05}. For more recent discussions of the observational situation see \citet{blake07}, \citet{padmanabhan07}, and \citet{percivaletal07a, percivaletal07b}.

The sound horizon at decoupling can be computed from relatively well-measured quantities by using relatively well-established physics. Consequently it is a standard ruler and can be used to trace the universe's expansion dynamics \citep[see, e.g.,][and references therein]{blake03, linder03, seo03, hu03}. A measurement of the BAO peak length scale at redshift $z$ fixes a combination of the angular diameter distance and Hubble parameter at that redshift. More precisely, what is determined \citep{eisensteinetal} is the distance

\begin{equation}
\label{eq:distancem}
D_V(z)=\left[(1+z)^2d_A^2(z)z/H(z)\right]^{1/3},
\end{equation}

\noindent
where $H(z)$ is the Hubble parameter and the angular diameter distance

\begin{equation}
\label{eq:angdis}
(1+z)d_A(z) = \begin{cases}
              H_0a_0R\sinh\left(\displaystyle\int_0^z dz'/\left(H(z')a_0R\right)\right) & \text{\ open model},\\
              H_0\displaystyle\int_0^z dz'/H(z') & \text{\ flat model}, \\
              H_0a_0R\sin\left(\displaystyle\int_0^z dz'/\left(H(z')a_0R\right)\right) & \text{\ closed model},
              \end{cases}
\end{equation}

\noindent
in the notation of \citet[][Chap.\ 13]{peeblesbook}. Here $a_0R$ is the radius of curvature of spatial hypersurfaces at fixed time and $H_0$ is the present value of the Hubble parameter. $D_V(z)$ depends on the cosmological parameters of the model, including those which describe dark energy, so we can constrain these parameters by comparing the predicted $D_V(z)$ to the measurements.

\section{Computation}
In this paper we study the $\phi$CDM model and compare our results with those derived in the standard $\Lambda$CDM model and the XCDM parametrization. In all three cases two parameters completely describe the background dynamics. For $\Lambda$CDM this pair is the nonrelativistic matter density parameter $\Omega_{\rm m}$ and the cosmological constant density parameter $\Omega_\Lambda$, both defined relative to the critical energy density today. In the $\Lambda$CDM model we study, the spatial curvature density parameter $\Omega_{\rm k}=1-\Omega_{\rm m}-\Omega_\Lambda$ need not vanish. For $\phi$CDM the model parameters are $\Omega_{\rm m}$ and a nonnegative constant $\alpha$ which characterizes the scalar field potential energy density $V(\phi)\varpropto\phi^{-\alpha}$. In the $\phi$CDM case we consider only spatially-flat models and $\phi$CDM at $\alpha=0$ is equivalent to $\Lambda$CDM with the same $\Omega_{\rm m}$ and $\Omega_{\rm k}=0$. The XCDM parametrization is characterized by $\Omega_{\rm m}$ and the negative equation of state parameter $\omega_{\rm x}=p/\rho$. Spatial curvature is also taken to be zero for the XCDM case and XCDM at $\omega_{\rm x}=-1$ is equivalent to $\Lambda$CDM with the same $\Omega_{\rm m}$ and $\Omega_{\rm k}=0$.

To compare theoretical predictions with observations we have to compute the angular diameter distance as a function of redshift for all three dark energy models. The angular diameter distance (equation~\ref{eq:angdis}) depends on the Hubble parameter as a function of redshift. The Hubble parameters in the $\Lambda$CDM and XCDM models are given by

\begin{equation}
H(z)=H_0\sqrt{\Omega_m(1+z)^3+\Omega_\Lambda+(1-\Omega_m-\Omega_\Lambda)(1+z)^2},
\end{equation}

\noindent
and

\begin{equation}
H(z)=H_0\sqrt{\Omega_m(1+z)^3+(1-\Omega_m)(1+z)^{3(1+\omega_{\rm x})}}.
\end{equation}

\noindent
For the $\phi$CDM model we consider $H(z)$ does not have an analytical expression. Instead one has to solve the coupled system of differential equations,

\begin{equation}
H(z)=H_0\sqrt{\Omega_m(1+z)^3+(\dot{\phi}^2+k\phi^{-\alpha}/G)/12},
\end{equation}

\begin{equation}
\ddot{\phi}+3\dot{\phi}-\frac{k\alpha}{2G}\phi^{-(\alpha+1)}=0,
\end{equation}

\noindent where $k$ is a constant. Figure\ 1 shows the residuals of
comoving distance calculated in $\phi$CDM and XCDM as a function of
redshift. The $\phi$CDM values are computed for $\Omega_{rm m}=0.3$
and $\alpha=2$, XCDM predictions are computed for the same value of
$\Omega_{ rm m}$ and three diferent values of $\omega_{\rm x}$:
$\omega_{\rm x}=-0.5$ (solid line), $\omega_{\rm x}=-1.0$ (dashed
line), and $\omega_{\rm x}=-2.0$ (dotted line). In general, the
theoretical lines do not reduce to each other for all redshifts for
any set of parameters $\alpha$ and $\omega_{\rm x}$ (other then
$\alpha=0$ which is the same as $\omega_{\rm x}=-1$). Figure\ 1
shows that for $\alpha=2$ the predictions of $\phi$CDM and XCDM
models already differ significantly..

We examine the constraints on the two cosmological parameters for each dark energy model from two measurements of the BAO peak. The first is from the BAO peak measured at $z=0.35$ in the correlation function of luminous red galaxies in the SDSS \citep{eisensteinetal}. This measurement results in $A(0.35)=0.469\pm0.017$ (one standard deviation error), where the dimensionless and $H_0$-independent function

\begin{equation}
A(z)=D_V(z)\frac{\sqrt{\Omega_{\rm m}H_0^2}}{z}
\end{equation}

\noindent
and $D_V(z)$ is the distance measure defined in equation (\ref{eq:distancem}). The measured value of $A(0.35)$ does not depend on the dark energy model and only weakly depends on the baryonic energy density \citep[see Sec.\ 4.5 in][]{eisensteinetal}. The measurement also has a weak dependence on parameters like the spectral index of primordial scalar energy density perturbations (the assumed value is $n=0.98$) and the sum of the neutrino masses, but this is not strong enough to have significant effect on the final result. To constrain cosmological model parameters in this case we perform a standard $\chi^2$ analysis.

The second BAO peak measurement we use is from the correlation function of galaxy samples drawn from the SDSS and 2dFGRS at two different redshifts, $z=0.2$ and $z=0.35$, as determined by \citet{percivaletal07a}.\footnote{This analysis includes the SDSS luminous red galaxies, so the \citet{percivaletal07a} and \citet{eisensteinetal} BAO peak measurements are not statistically independent. There are a number of ways to use the \citet{percivaletal07a} BAO peak measurements to constrain cosmological parameters. Here we use their $S_{\rm k}/D_{\rm V}$ method to compute constraints.} This measurement gives the correlated values $r_s/D_V(0.2)=0.1980\pm0.0058$ and $r_s/D_V(0.35)=0.1094\pm0.0033$ (one standard deviation errors), where $r_s$ is the comoving sound horizon at recombination, equation \ (8) of \citet{percivaletal07a}. These two measurements are correlated, with the inverse of the correlation matrix given by

\begin{equation}
V^{-1}=\left( \begin{array}{c c}
35059& -24031\\
-24031& 108300\\
\end{array}
\right).
\nonumber
\end{equation}

\noindent
To compute $r_s$ we first compute the angular diameter distance to the surface of last scattering, $d_A(1089)$. We then use the WMAP measurement of the apparent acoustic horizon angle in the CMB anisotropy data \citep{spergel07} to determine the sound horizon $r_s=[(1+z)d_A(z)]|_{z=1089}\times 0.0104$ (where we ignore the WMAP measurement uncertainty and assume that $r_s$ is known perfectly). The use of the WMAP prior on the apparent acoustic horizon angle results in very tight constraints on the spatial curvature. When this measurement is not used, the \citet{percivaletal07a} measurements alone can not tightly constrain the dark energy parameters \citep[see the shaded areas in Fig.\ 12 of][]{percivaletal07a}.

To constrain cosmological parameters in this case we follow \citet{percivaletal07a} and first compute
\begin{equation}
X(\Omega_{\rm m}, \alpha)=\left( \begin{array}{c}
 r_s/D_V(0.2,\Omega_{\rm m},\alpha)-0.1980\\
r_s/D_V(0.35,\Omega_{\rm m},\alpha)-0.1094\\
\end{array} \right),
\end{equation}

\noindent
where for definiteness we consider the $\phi$CDM model. We then compute the $\chi^2$ function

\begin{equation}
\chi^2(\Omega_{\rm m}, \alpha)=X^{-1}V^{-1}X.
\end{equation}
\noindent
and the likelihood function

\begin{equation}
L(\Omega_{\rm m}, \alpha)={\rm exp}(-\chi^2(\Omega_{\rm m}, \alpha)/2).
\end{equation}

For both the \citet{eisensteinetal} and the \citet{percivaletal07a} measurements and for all the models we consider, $\Lambda$CDM, $\phi$CDM, and XCDM, $\chi^2$ is a function of two parameters, either $(\Omega_{\rm m},\Omega_\Lambda)$, $(\Omega_{\rm m}, \alpha)$, or $(\Omega_{\rm m},\omega_{\rm x})$. To define 1$\sigma$, 2$\sigma$, and 3$\sigma$ confidence level contours in these two-dimensional parameter spaces, we pick sets of points with $\chi^2$ values larger than the minimum $\chi^2$ value by $2.3$, $6.17$, and $11.8$, respectively.

Figures\ 2,\ 3, and\ 4 show constraints on $\Lambda$CDM, XCDM, and $\phi$CDM from the \citet[dashed lines]{eisensteinetal} and \citet[solid lines]{percivaletal07a} data. Parts of the contours are not smooth because of computational noise. The BAO peak contours in these figures show that the measurement essentially constrains only one free parameter. When BAO peak measurements at other redshifts become available in the future, BAO data should then constrain both cosmological parameters.

For now, to break this degeneracy and constrain both free parameters we use these BAO results together with constraints from galaxy cluster gas mass fraction versus redshift data \citep{samushia08}. The new galaxy cluster gas mass fraction data \citep{allen08} gives the ratio of X-ray emitting hot baryonic gas mass to total gravitational mass for 42 hot, dynamically relaxed galaxy clusters in a redshift range from $z=0.05$ to $z=1.1$. Since the gas mass fraction of these relaxed clusters is expected to be independent of redshift these measurements can be used to constrain cosmological model parameters.

In a given cosmological model, the predicted cluster gas mass fraction also depends on the value of the Hubble constant and the density of baryonic matter. We treat these as ``nuisance" parameters, assume prior probability distribution functions for them, and marginalize over them to derive the probability distribution function for the pairs of cosmological parameters of interest \citep[see, e.g.,][]{ganga97}. Since there still is some uncertainty in the values of these parameters we use two sets of Gaussian priors in our computations. One is the set $h=0.73\pm0.03$ and $\Omega_{\rm b}h^2=0.0223\pm0.0008$ from WMAP data \citep{spergel07}, the second is $h=0.68\pm0.04$ \citep{got01,che03c} and $\Omega_{\rm b}h^2=0.0205\pm0.0018$ \citep{fie06}, all one standard deviation errors.\footnote{The priors from the WMAP data that we use are derived for the best fit $\Lambda$CDM model. For other models and different parameter values the estimates on $h$ and $\Omega_{\rm b}$ will be slightly different. Since the joint likelihood function does not strongly depend on $h$ and $\Omega_{\rm b}$ priors (for reasonable ranges of the priors) this does not lead to big differences in the final result. If one wants to be more rigorous, slightly broader priors from the HST Key Project \citep{freedman01} and big bang nucleosynthesis \citep{kirkman03} can be used.} Confidence level contours derived from \citet{allen08} cluster gas mass fraction data are shown in Figs.\ 2,\ 3, and\ 4 as two sets of dotted lines corresponding to the two sets of priors for the Hubble constant and baryonic matter mass density.

Since the gas mass fraction and BAO peak measurements are statistically independent we define the joint $\chi^2$ function by adding together the individual $\chi^2$s. The resulting joint constraints are shown in Figs.\ 5, 6, and 7.

\section{Results and discussion}

The \citet{eisensteinetal} BAO peak measurement has been used in conjunction with other data to place constraints on various cosmological models \citep[see, e.g.,][]{alam06, nesseris07, movahed07, zhangwu07, wright07, shafieloo07}. The more recent \citet{percivaletal07a} data has also been used for this purpose \citep{ishida08, lazkoz07a}.

Constraints from BAO peak measurements and galaxy cluster gas mass
fraction data are shown in Figs.\ 2, 3, and 4. The solid line contours
in Figs.\ 2 and 3 show the constraints on $\Lambda$CDM and XCDM derived
from the \citet{percivaletal07a} BAO data and are comparable to those
shown with dashed lines in Fig.\ 12 in their paper. The dashed contours
in Fig.\ 3 are comparable to those shown in Fig.\ 11 in
\citet{eisensteinetal}. \citet{eisensteinetal} do not show
contours for $\Lambda$CDM (see our Fig.\ 2) and the BAO contours we show
in Fig.\ 4 have not previously been presented. Figure\ 2 shows that the
\citet{percivaletal07a} constraints, which make use of the WMAP
measurement of the apparent acoustic horizon angle, constrain the sum of parameters
$\Omega_\Lambda$ and $\Omega_{\rm m}$ to be very close to one
($\Omega_{\rm k}=1-\Omega_{\rm m}-\Omega_\Lambda \approxeq 0$) and
favor a close to spatially flat model if dark energy is time
independent. The spatial curvature is constrained so well mainly because we use
the WMAP measurement of the apparent acoustic peak angle. BAO measurements by
themselves can not effectively constrain dark energy parameters very well
\citep[see shaded areas in Fig.\ 12 of][]{percivaletal07a}.
In spatially-flat models BAO peak measurements put
tight constraints on the $\Omega_{\rm m}$ parameter; they do not
well constrain the ``orthogonal'' cosmological parameter $\Omega_\Lambda$ and in particular
they allow dark energy to vary in time (see Figs.\ 3 and 4).

The BAO constraints are significantly tighter than the
Hubble parameter versus redshift data ones \citep[see,
e.g.,][]{samushia07} and the strong gravitational lensing ones
\citep[see, e.g.,][]{cha04}. They are, in general, about as
constraining as the SNIa results and constrain roughly the same linear combination
of cosmological parameters \citep[see, e.g.,][]{wilson06}. In Figs.\ 2 and\ 3 the best fit
values from \citet{percivaletal07a} measurement and WMAP prior are more then
3$\sigma$ away from the best fit of the cluster gas mass fraction constraints.
This is most probably due to unknown systematic errors in one or both of the
measurements or an effect of poor statistics and should change when more and better data are available.

 The joint BAO peak and cluster gas mass fraction constraints are shown
 in Figs.\ 5, 6, and 7. They are fairly restrictive and favor a
 spatially-flat $\Lambda$CDM model with $\Omega_{\rm m} \sim $ 0.25 and $\alpha < 0.5$ on $1\sigma$ confidence level.
 Since the predictions of $\phi$CDM for a very small value of the $\alpha$ parameter are very
 close to the predictions of the spatially-flat $\Lambda$CDM model, current observational tests
 are unable to discriminate between a time-independent cosmological constant and a slowly varying
 scalar field with $\alpha$ of order 1. All three models considered here give about the
 same $\chi^2 \sim 52$ for 41 degrees of freedom and there is no reason to
 favor one model over another based on Bayesian statistics. The
 constraints from the joint analysis on all three dark energy models are comparable to the constraints
 derived from a joint analysis \citep{wilson06} of earlier SNIa data \citep{riess04} and earlier cluster gas mass
 fraction data \citep{allen04}. Constraints on $\alpha$ derived from the joint analysis are stronger than the results quoted in previously published papers
 (see Table\ 1)
 The joint BAO peak and gas mass fraction data constraints on $\Lambda$CDM and XCDM derived here are a
 little weaker than those derived from BAO peak and more recent SNIa
 \citep{astier06} data, see Figs.\ 13 of \citet{percivaletal07a}. In the joint analysis done here the uncertainties on $h$ and $\Omega_{\rm b}h^2$ play a less significant role than they do in the cluster gas mass fraction analysis, i.e., the contours for the two prior sets are closer to each other in Figs.\ 5,\ 6, and\ 7, than in Figs.\ 2,\ 3, and\ 4. The contours in Figs.\ 5 and\ 6 are in agreement with tighter joint results from other data sets considered by \citet{wangetal07}.

From Fig.\ 1 it is clear that for large values of $\alpha$ $\phi$CDM
and XCDM models predict different cosmological evolution. For small
values of $\alpha$, however, if parameters are chosen appropriately,
different dark energy models will at low redshifts predict very
similar background evolution. Because of that, low redshift distance
measurements have to be complemented with high redshift CMB and
large scale structure measurements to discriminate between dark
energy models. Better quality BAO peak data at a number of redshifts
and more gas mass fraction measurements, along with tighter priors
on nuisance parameters like the Hubble parameter and the density of
baryonic matter, will allow for tighter constraints on dark energy
parameters and could soon either detect a time dependence in dark
energy or constrain it to a very small value.

\acknowledgements
We acknowledge helpful discussions with M. Vogeley and support from DOE grant DE-FG03-99EP41093 and INTAS grant 061000017-9258.

\clearpage

\begin{deluxetable}{l l l}
\tablehead{\colhead{3$\sigma$ constraints on $\alpha$} & \colhead{Observational test(s) used} & \colhead{reference}}
\tablecaption{Constraints on the $\phi$CDM $\alpha$ parameter from different observational tests.}
\startdata
$\alpha$ \text{\ not well constrained}& \text{SNIa} & \citet{podariu00} \\
$\alpha$ \text{\ not well constrained}& \text{Radio galaxies} & \citet{pod03} \\
$\alpha$ \text{\ not well constrained}& \text{Gravitational lensing} & \citet{cha04} \\
$\alpha<6.5$ & \text{Galaxy clusters} & \citet{che04} \\
$\alpha<5$ & \text{Radio galaxies \& SNIa ``Gold'' data set} & \citet{wilson06} \\
$\alpha<5$ & \text{Radio galaxies \& SNIa} & \citet{dal07} \\
$\alpha<4.5$ & \text{Galaxy clusters} & \citet{samushia08} \\
$\alpha<3.5$ & \text{Galaxy clusters \& BAO} & \text{This work} \\
\enddata
\end{deluxetable}

\begin{figure}
\includegraphics[height=6.5in, width=6.5in, trim=0in 0in 0in 0in, clip, scale=1]{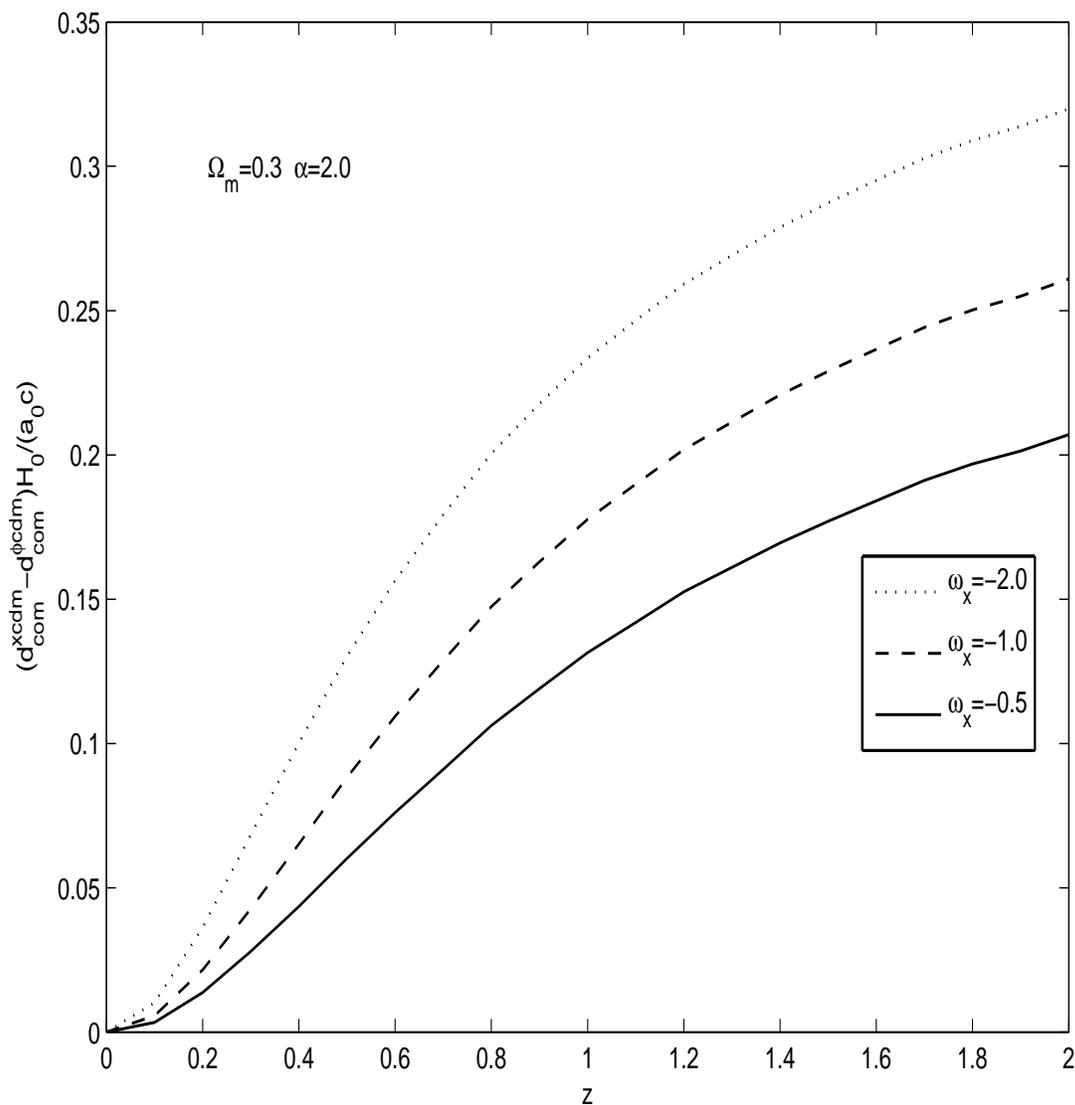}
\caption{Residuals of comoving distance calculated in the $\phi$CDM model with $\Omega_{\rm m}=0.3$ and $\alpha=2$ and predictions of XCDM parameterization with $\omega_{\rm x}=-0.5$ (solid line), $\omega_{\rm x}=-1.0$ (dashed line), and $\omega_{\rm x}=-2.0$ (dotted line). Comoving distance is normalized to $a_0c/H_0$.}
\end{figure}

\begin{figure}
\includegraphics[height=6.5in, width=6.5in, trim=0in 0in 0in 0in, clip, scale=1]{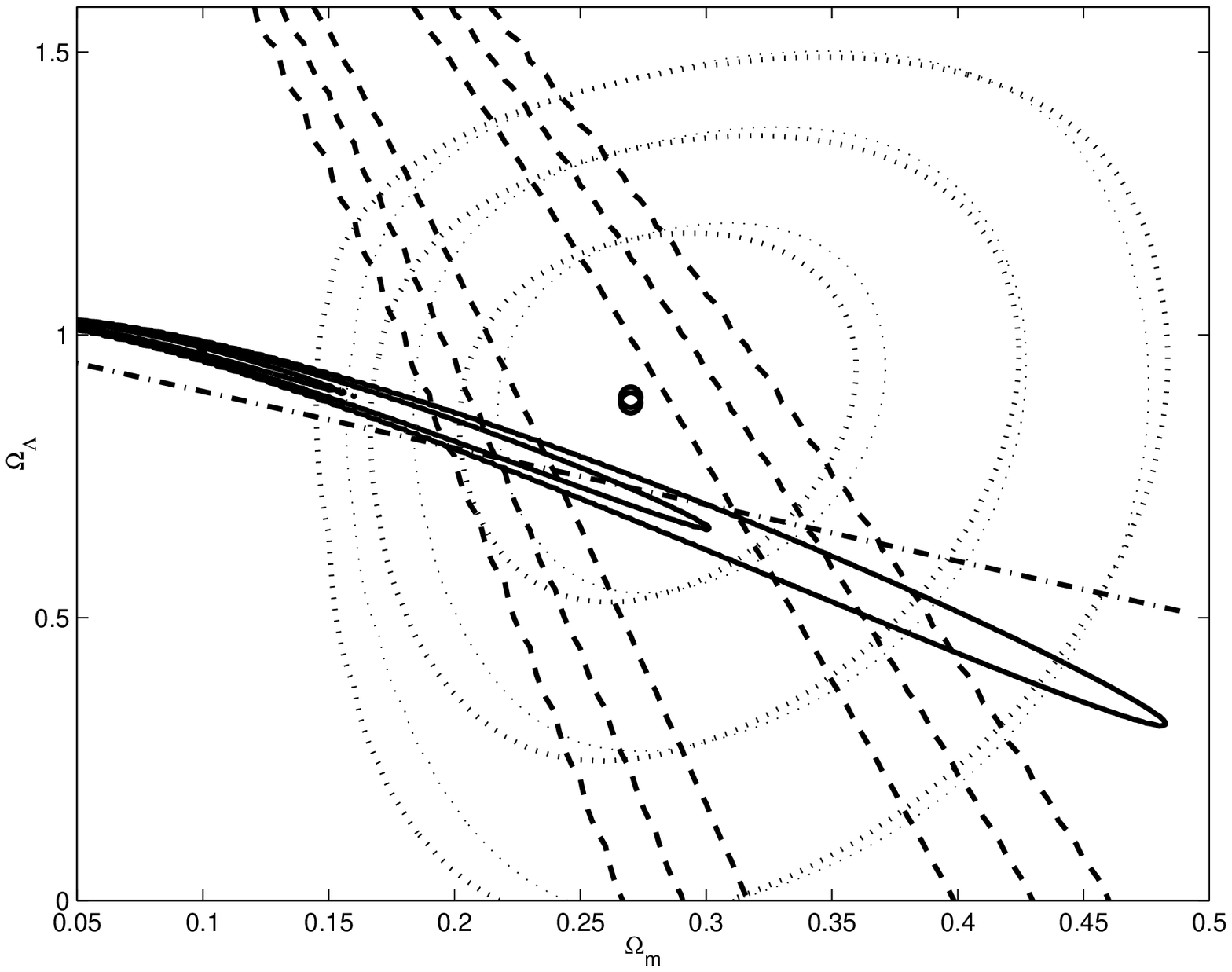}
\caption{1, 2, and 3$\sigma$ confidence level contours for the $\Lambda$CDM model.
Solid lines are the constraints derived from \citet{percivaletal07a} BAO measurements together with WMAP measurement of acoustic horizon angle
(the $\times$ for the best fit value near $\Omega_{\rm m}$ is obscured) and dashed
lines are those from the \citet{eisensteinetal} BAO measurement. The two sets of dotted
lines are the constraints derived from galaxy cluster gas mass fraction data \citep{samushia08}
($\circ$ for the best fit value); the thick dotted lines are derived using the WMAP priors
for $h$ and $\Omega_{\rm b}h^2$ while the thin dotted lines are for the alternate priors (see text).
The dot-dashed line corresponds to spatially-flat $\Lambda$CDM models.}
\end{figure}

\begin{figure}
\includegraphics[height=6.5in, width=6.5in, trim=0in 0in 0in 0in, clip, scale=1]{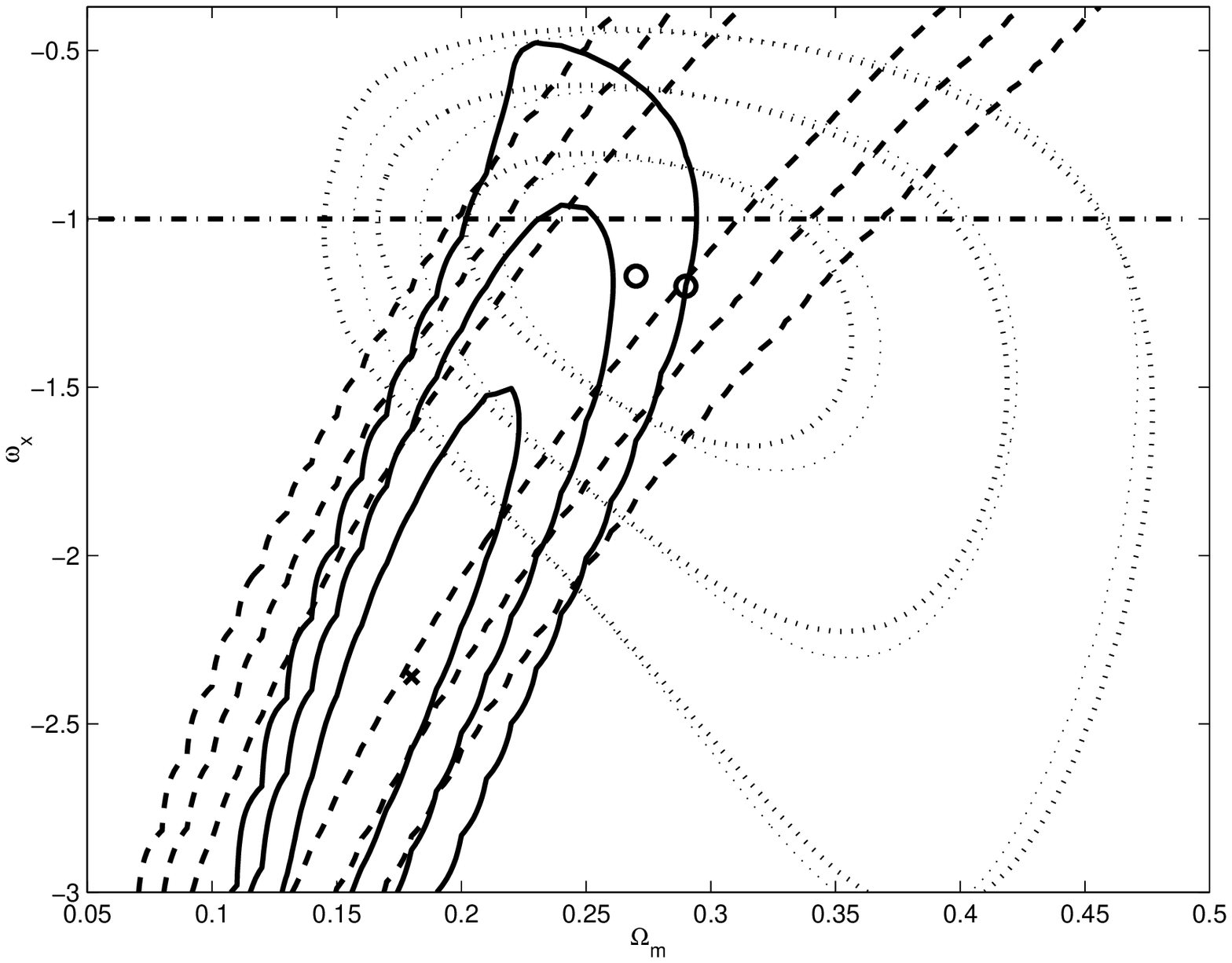}
\caption{1, 2, and 3$\sigma$ confidence level contours for the XCDM parametrization. Solid lines are the constraints derived from \citet{percivaletal07a} BAO measurements together with WMAP measurement of acoustic horizon angle ($\times$ for the best fit value) and dashed lines are those from the \citet{eisensteinetal} BAO measurement. The two sets of dotted lines are the constraints derived from galaxy cluster gas mass fraction data \citep{samushia08} ($\circ$ for the best fit value); the thick dotted lines are derived using WMAP priors for $h$ and $\Omega_{\rm b}h^2$ while the thin dotted lines are for the alternate priors (see text). The horizontal dot-dashed line corresponds to spatially-flat $\Lambda$CDM models.}
\end{figure}

\begin{figure}
\includegraphics[height=6.5in, width=6.5in, trim=0in 0in 0in 0in, clip, scale=1]{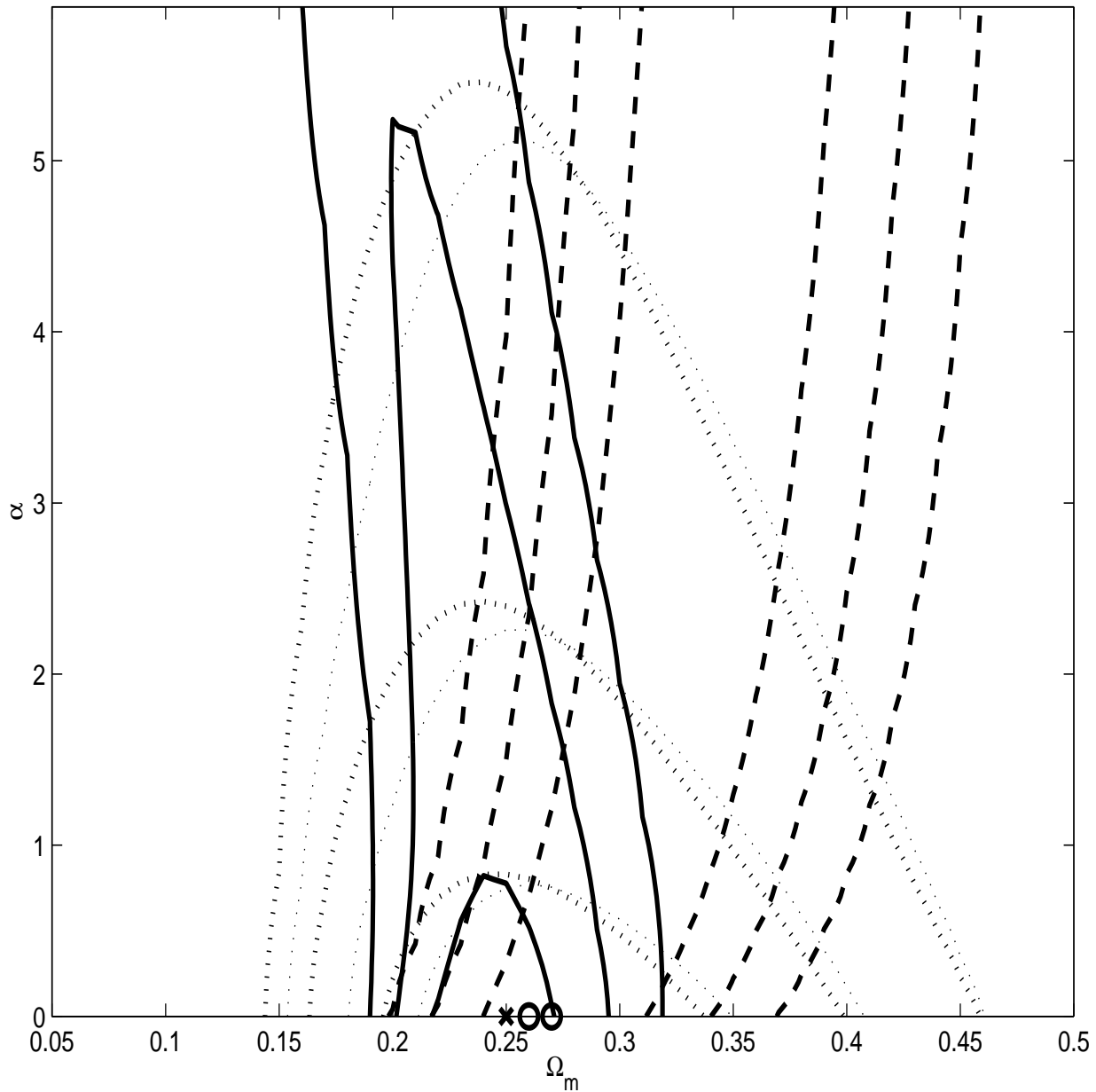}
\caption{1, 2, and 3$\sigma$ confidence level contours for the $\phi$CDM model. Solid lines are the constraints derived from \citet{percivaletal07a} BAO measurements together with WMAP measurement of acoustic horizon angle ($\circ$ for the best fit value) and dashed lines are those from the \citet{eisensteinetal} BAO measurement. The two sets of dotted lines are the constraints derived from galaxy cluster gas mass fraction data \citep{samushia08} ($\times$ for the best fit value); the thick dotted lines are derived using WMAP priors for $h$ and $\Omega_{\rm b}h^2$ while the thin dotted lines are for the alternate priors (see text). The $\alpha=0$ axis corresponds to spatially-flat $\Lambda$CDM models.}
\end{figure}

\begin{figure}
\includegraphics[height=6.5in, width=6.5in, trim=0in 0in 0in 0in, clip, scale=1]{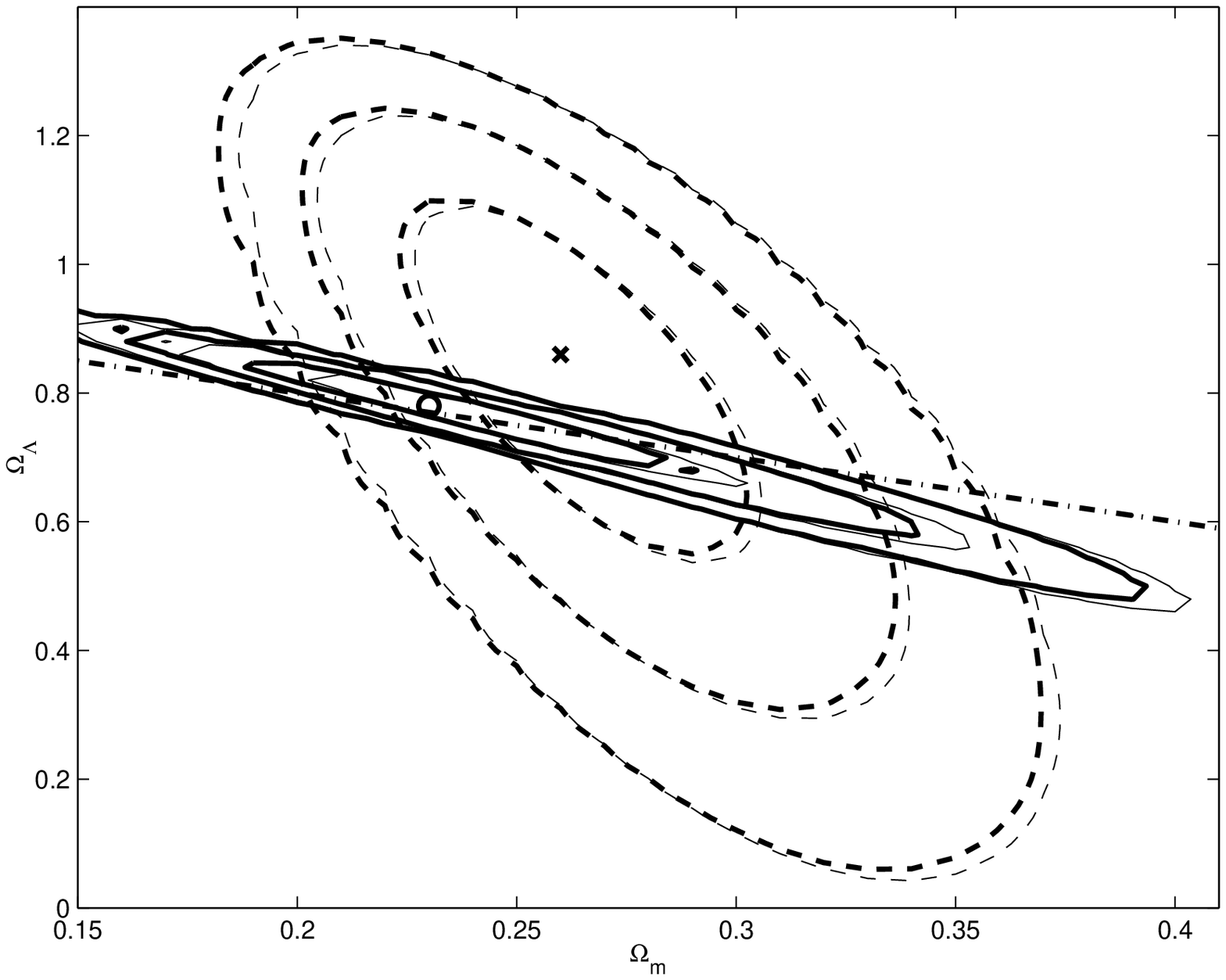}
\caption{1, 2, and 3$\sigma$ confidence level contours for the
    $\Lambda$CDM model. Solid lines are the joint constraints
        derived from \citet{percivaletal07a} BAO measurements together with WMAP measurement of acoustic horizon angle
        and galaxy cluster gas mass fraction data ($\circ$
                shows the best fit value with an acceptable $\chi^2\simeq 57$ for 42 degrees of freedom); dashed lines
        are the corresponding joint constraints using the
        \citet{eisensteinetal} BAO measurement ($\times$ shows
                the best fit value with an acceptable $\chi^2\simeq 52$ for 41 degrees of freedom). Thick lines are
        derived using the WMAP priors for $h$ and $\Omega_{\rm
            b}h^2$ while thin lines are for the alternate
            priors. The joint best fit values for the two
            prior sets overlap. The dot-dashed line
            corresponds to spatially-flat $\Lambda$CDM
            models. The $\Omega_{\rm m}$ and
            $\Omega_{\Lambda}$ ranges shown here are smaller
            than those shown in Fig.\ 2.}
\end{figure}

\begin{figure}
\includegraphics[height=6.5in, width=6.5in, trim=0in 0in 0in 0in, clip, scale=1]{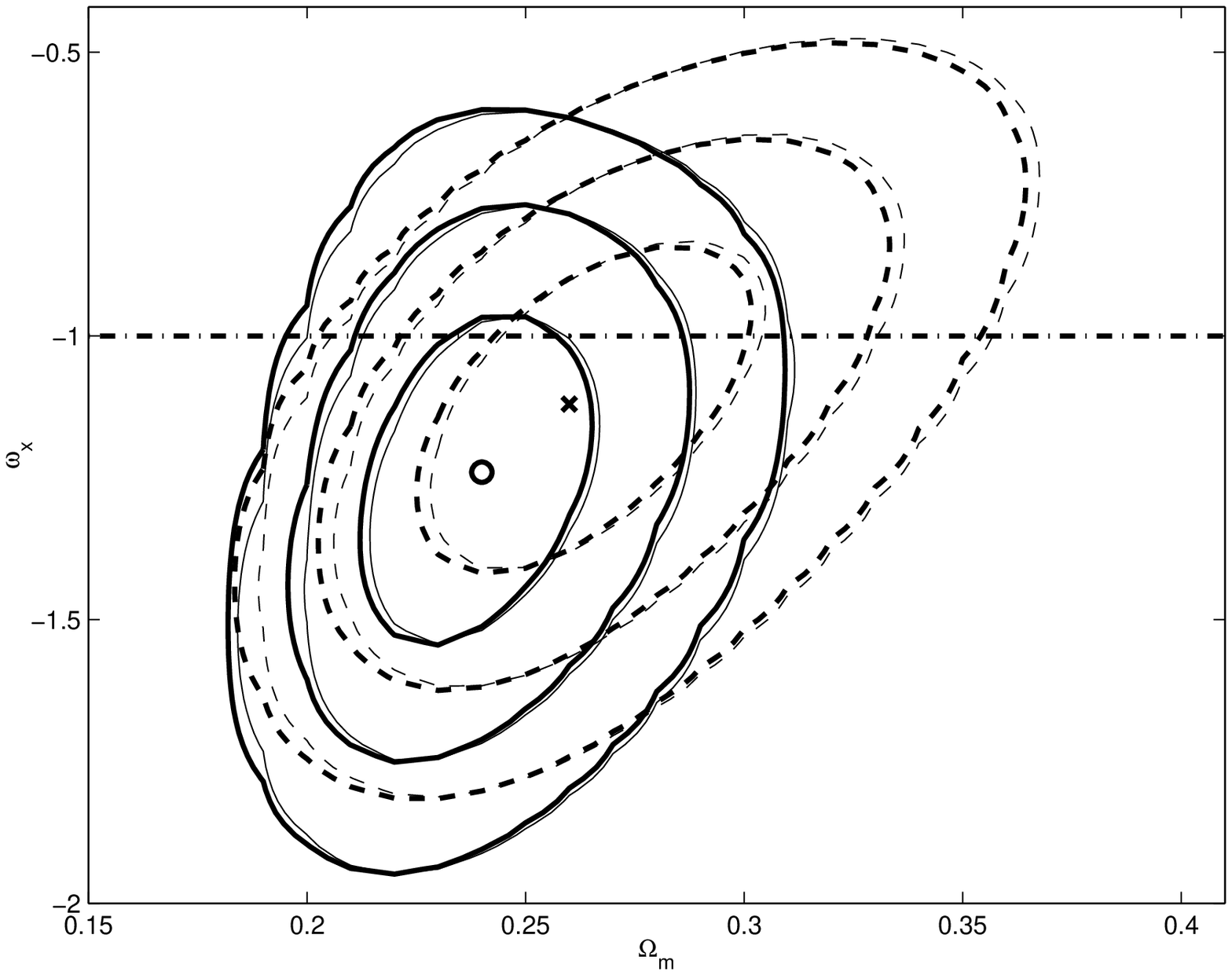}
\caption{1, 2, and 3$\sigma$ confidence level contours for the XCDM
    parametrization. Solid lines are the joint constraints derived
        from \citet{percivaletal07a} BAO measurements together with WMAP measurement of acoustic horizon angle and galaxy
        cluster gas mass fraction data ($\circ$ shows the best
                fit value with an acceptable $\chi^2\simeq 56.5$ for 42 degrees of freedom); dashed lines are the
        corresponding joint constraints using the
        \citet{eisensteinetal} BAO measurement ($\times$ shows
                the best fit value with an acceptable $\chi^2\simeq 52$ for 41 degrees of freedom). Thick lines are
        derived using the WMAP priors for $h$ and $\Omega_{\rm
            b}h^2$ while thin lines are for the alternate
            priors. The joint best fit values for the two
            prior sets overlap. The horizontal dot-dashed
            line corresponds to spatially-flat $\Lambda$CDM
            models. The $\Omega_{\rm m}$ and $\omega_{\rm
                x}$ ranges shown here are smaller than
                those shown in Fig.\ 3.}
\end{figure}

\begin{figure}
\includegraphics[height=6.5in, width=6.5in, trim=0in 0in 0in 0in, clip, scale=1]{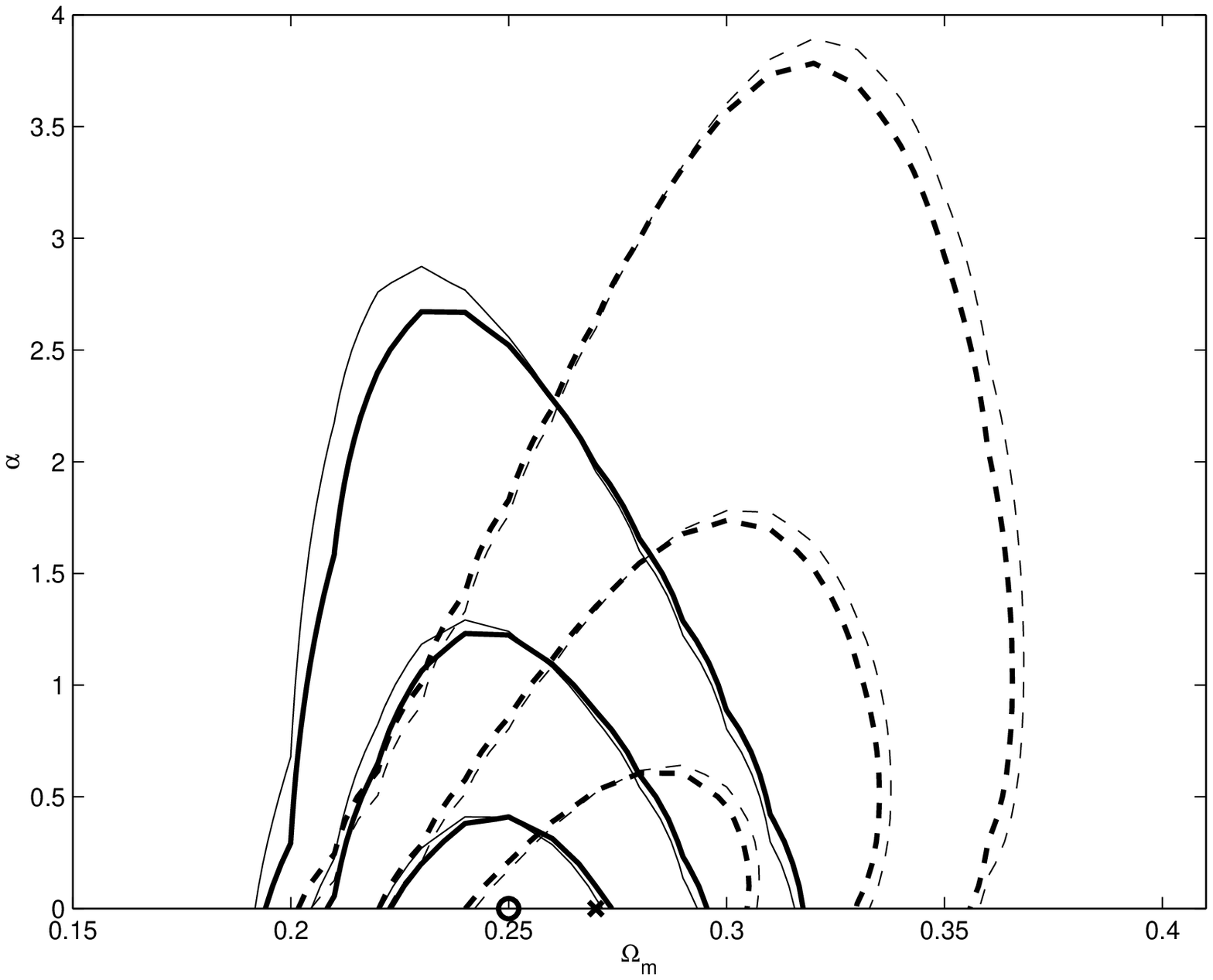}
\caption{1, 2, and 3$\sigma$ confidence level contours for the $\phi$CDM
    model. Solid lines are the joint constraints derived from
        \citet{percivaletal07a} BAO measurements together with WMAP measurement of acoustic horizon angle and galaxy
        cluster gas mass fraction data ($\circ$ shows the best
                fit value with an acceptable $\chi^2\simeq 58$ for 42 degrees of freedom); dashed lines are the
        corresponding joint constraints using the
        \citet{eisensteinetal} BAO measurement ($\times$ shows
                the best fit value with an acceptable $\chi^2\simeq 52$ for 41 degrees of freedom). Thick lines are
        derived using the WMAP priors for $h$ and $\Omega_{\rm
            b}h^2$ while thin lines are for the alternate
            priors. The joint best fit values for the two
            prior sets overlap. The $\alpha=0$ axis
            corresponds to spatially-flat $\Lambda$CDM
            models. The $\Omega_{\rm m}$ and $\alpha$ ranges
            shown here are smaller than those shown in Fig.\ 4.}
\end{figure}

\end{document}